\documentclass{article}
\usepackage{amssymb}
\begin{document}
\date{\today}

\title {Why does gravitational radiation produce vorticity?}
\author{L Herrera$^1$, W Barreto$^2$, J Carot$^3$, A Di Prisco$^1$\\
$^1$ Escuela de F\'{\i}sica, Facultad de Ciencias,\\
Universidad Central de Venezuela, Caracas, Venezuela.\\
$^2$ Centro de F\'\i sica Fundamental,
Facultad de Ciencias,\\ Universidad de los Andes, M\'erida, Venezuela.\\
$^3$ Departament de  F\'{\i}sica, Universitat Illes Balears,\\
E-07122 Palma de Mallorca, Spain.}
\maketitle
\begin{abstract}
We calculate  the vorticity of world--lines of observers at rest in a
Bondi--Sachs frame, produced by gravitational radiation, in a general Sachs
metric. We claim that such an effect is
related to  the super--Poynting
vector, in a similar way as the existence of  the
electromagnetic Poynting vector is
related to the vorticity in stationary electrovacum spacetimes.

\end{abstract}

\section{Introduction}
The theoretical description and the experimental observation of
gravitational radiation are among the most relevant challenges
confronting general relativity.

A great deal of work has been done so far in order to provide
a consistent framework for the study of such phenomenon. Also,
important collaboration efforts have been carried on, and are
now under consideration, to put in evidence gravitational waves.

Therefore it is clear that any specific phenomenon caused by gravitational
radiation and which might be observed, could in principle lead to the
detection of such elusive waves and are
of utmost relevance.

In this respect, particular attention  deserves the link  between  gravitational
radiation and vorticity of world--lines of observers at rest in a Bondi
space--time \cite{HHP, Valiente, HSC}.
Specifically, it has been shown that the leading term in the vorticity (in
an expansion of powers of $1/r$) is expressed through the news function in
such a way that it will vanish if
and only if there  is no news (no
radiation). This suggests the possibility
of detecting gravitational waves
by means of gyroscopes \cite{HHP, Felice}.

The issue we want to address
here is: What is the mechanism by means of
which gravitational radiation
produces vorticity?

To answer to such a question  it
is worth recalling a
result obtained by Bonnor \cite{bonleter} concerning
the dragging of
inertial frames by a
charged magnetic dipole. To explain the appearance of
vorticity in such
space--times, Bonnor notices that the corresponding
electromagnetic
Poynting vector has a non--vanishing
component, describing
a flow of electromagnetic energy round in circles
where frame--dragging
occurs \cite{Feymann}. He then suggests that such a
flow of energy affects
inertial frames
by producing vorticity of congruences of particles,
relative to the compass
of inertia. This conjecture has been recently
confirmed to be valid for
general axially symmetric stationary
electrovacumm metrics \cite{HCol}.

One is then tempted to  speculate that
a similar mechanism is working in
the case of gravitational radiation, i.e.
a flow of
gravitational ``energy''  would
produce
vorticity of congruence of observers. For
testing such a conjecture
we have available a tensor quantity which allows
to define a
covariant
(super)--energy density and a (super)--Poynting vector, namely
the
Bel--Robinson tensor \cite{Bel}.

When  the  super--Poynting
vector ($P^{\mu}$), based on the
Bel--Robinson tensor, as defined in
\cite{Basset},  is calculated for the
Bondi metric \cite{Boal}, we obtain
that the contravariant azimuthal component ($P^{\phi}$) of such vector
vanishes\cite{HSC}, as expected from the the reflection symmetry
of the Bondi metric, which, intuitively, seems to
be incompatible with the presence of a circular flow of energy in the $\phi$ direction. However, the vorticity vector, which is orthogonal to the plane of rotation,
has in the Bondi spacetime only one non--vanishing contravariant component ( $\phi$ ). Implying thereby that the plane of the associated rotation is  orthogonal to the
$\phi$ direction. Therefore, it is not the  $\phi$ component of $P^{\mu}$ ( as  mistakenly stated in \cite{HSC}) the possible source of vorticity, but the $\theta$
component, the cause for such an effect, in the Bondi case.

In order to strenght further the case for the super-Poynting vector as the cause of the mentioned vorticity, we shall consider here  the general radiative metric
without axial and reflection symmetry
\cite{Sachs, Burg}. In this latter case we
shall obtain a non--vanishing
$P^{\phi}$, which we will identify as
the
cause of the  $\theta$ component of the vorticity vector. The discussion on these results
is
presented in the last section. However, before considering the
general
(Sachs) case it is instructive to review
very briefly the basic
ideas of the Bondi formalism in the reflection
symmetric case.

\section{The Bondi's formalism}
The general form of an axially and
reflection symmetric asymptotically flat
metric given by Bondi is
\cite{Boal}
\begin{eqnarray}
ds^2 & = & \left(\frac{V}{r} e^{2\beta} - U^2
r^2 e^{2\gamma}\right) du^2
+ 2 e^{2\beta} du dr \nonumber \\
& + & 2 U r^2
e^{2\gamma} du d\theta
- r^2 \left(e^{2 \gamma} d\theta^2 + e^{-2\gamma}
\sin^2{\theta}
d\phi^2\right)
\label{Bm}
\end{eqnarray}
where $V, \beta, U$
and $\gamma$ are functions of
$u, r$ and $\theta$.

We number the
coordinates $x^{0,1,2,3} = u, r, \theta, \phi$ respectively.
$u$ is a
timelike coordinate such that $u=constant$ defines a null surface.
In flat
spacetime this surface coincides with the null light cone
open to the
future. $r$ is a null coordinate ($g_{rr}=0$) and $\theta$ and
$\phi$ are
two angle coordinates (see \cite{Boal} for details).

The four metric
functions are assumed to be expanded in series of $1/r$,
then using the
field equations Bondi gets

\begin{equation}
\gamma = c r^{-1} + \left(C -
\frac{1}{6} c^3\right) r^{-3}
+
...
\label{ga}
\end{equation}
\begin{equation}
U = - \left(c_\theta + 2 c
\cot{\theta}\right) r^{-2} + \left[2
N+3cc_{\theta}+4c^2
\cot{\theta}\right]r^{-3}...
\label{U}
\end{equation}
\begin{eqnarray}
V &
= & r - 2 M\nonumber \\
& - & \left( N_\theta + N \cot{\theta}
-
c_{\theta}^{2} - 4 c c_{\theta} \cot{\theta} -
\frac{1}{2} c^2 (1 + 8
\cot^2{\theta})\right) r^{-1} +
...
\label{V}
\end{eqnarray}
\begin{equation}
\beta = - \frac{1}{4} c^2
r^{-2} + ...
\label{be}
\end{equation}
where $c$, $C$, $N$ and $M$ are
functions of $u$ and $\theta$, letters as
subscripts denote derivatives.

Bondi shows that field equations allow to
determine the $u$-derivatives of all
functions in $\beta$ and $\gamma$ with
the exceptions of news
functions
$c_u$.

Thus the main conclusions emerging from  Bondi's
approach may be summarized as follows.

 If $\gamma, M$ and $N$ are known for some
$u=a$(constant) and
$c_u$ (the news function) is known for all $u$ in the
interval
$a \leq u \leq b$,
then the system is fully determined in that
interval. In other words,
whatever happens at the source, leading to
changes in the field,
it can only do so by affecting $c_u$ and viceversa.
In the
light of this comment the relationship between news function
and the
occurrence of radiation becomes clear.

Now, for an observer at rest in the frame of
(\ref{Bm}), the four-velocity
vector has
components $u^{\alpha} =A^{-1}\delta^{\alpha}_{u}$,
with
\begin{equation}
A \equiv
\left(\frac{V}{r} e^{2\beta} - U^2 r^2
e^{2\gamma}\right)^{1/2}.
\label{A}
\end{equation}

Then, it can be shown
that for such an observer the vorticity
vector
\begin{equation}
\omega^\alpha=\frac{1}{2\sqrt{-g}}\eta^{\alpha\eta\iota\lambda}u_\eta
u_{\iota,\lambda},\qquad \eta_{\alpha \beta \gamma \delta} \equiv
\sqrt{-g} \;\;\epsilon_{\alpha \beta \gamma \delta}\label{vv}
\end{equation}
 where $\eta_{\alpha\beta\gamma\delta}=
+1$ for $\alpha, \beta, \gamma,
\delta$ in
even order, $-1$ for $\alpha,
\beta, \gamma, \delta$ in
odd order and $0$ otherwise,
 may be
written as
(see \cite{HHP} for details)
\begin{equation}
\omega^\alpha = \left(0, 0,
0,
\omega^{\phi}\right)
\label{oma}
\end{equation}
giving for the absolute value of $\omega^{\alpha}$ (keeping only the leading term)

\begin{eqnarray}
\Omega  = -\frac{1}{2r} ( c_{u \theta}+2 c_u \cot
\theta).
\label{Om2}
\end{eqnarray}

Therefore, up to order $1/r$, a
gyroscope at rest in (\ref{Bm}) will
precess as long as the system radiates
($c_{u} \not= 0$). 

Next,
the  super--Poynting vector  based on the Bel--Robinson tensor, as
defined
in Maartens and Basset \cite{Basset},
is
\begin{equation}
P_{\alpha}=\eta_{\alpha \beta \gamma
\delta}E^{\beta}_{\rho}H^{\gamma
\rho}u^{\delta},
\label{p1}
\end{equation}

where
$E_{\mu\nu}$ and $H_{\mu\nu}$, are the electric and magnetic parts of
Weyl
tensor, respectively, formed from Weyl tensor $C_{\alpha
\beta \gamma
\delta}$ and its dual
$\tilde C_{\alpha \beta \gamma \delta}$ by
contraction with the four
velocity vector given
by
\begin{equation}
E_{\alpha \beta}=C_{\alpha \gamma \beta
\delta}u^{\gamma}u^{\delta}
\label{electric}
\end{equation}
\begin{equation}
H_{\alpha \beta}=\tilde C_{\alpha \gamma \beta
\delta}u^{\gamma}u^{\delta}=
\frac{1}{2}\eta_{\alpha \gamma \epsilon
\delta} C^{\epsilon
\delta}_{\quad \beta \rho}
u^{\gamma}
u^{\rho},
\label{magnetic}
\end{equation}

The electric and
magnetic parts of Weyl tensor have been explicitly
calculated for the Bondi
metric in \cite{HSC}, using those expressions the
reader can easily verify
that in this case
$P^{\phi}=0$. As mentioned before, this is to be expected due to the reflection symmetry of the metric.

Indeed,  rotation singles out a specific direction of time, which is at variance with the equivalence of both directions of time, implicit in reflection symmetry. This is the
reason why the Bondi metric in the time independent limit becomes the static Weyl metric and not the general stationary spacetime.

On the other hand, since the associated rotation in this case,
takes place on the plane orthogonal to the $\phi$ direction, it is $P^{\theta}$  the quantity to be associated with vorticity. Indeed, the vorticity vector describes the
rate of rotation,  on its orthogonal plane, with respect to proper time of the set of neighboring particles, relative to the local compass of inertia \cite{Rindler}.  As it can be 
seen from expressions (\ref{sp1}--\ref{sp3}) below, in the Bondi limit, such a component is not vanishing.

Let us now analyze the general case.

\section{The general radiative metric}

Shortly after the publication of the seminal
paper by Bondi and coworkers,
Sachs \cite{Sachs} presented a generalization
of the Bondi formalism
relaxing the conditions of axial and
reflection
symmetry. In this  case the line element reads (we have found
more
convenient to follow the notation given in \cite{Burg, Vickers} which
is slightly different from the
original Sachs
paper)

\begin{eqnarray}
ds^2=(Vr^{-1}e^{2\beta}-r^2 e^{2\gamma}U^2\cosh
2\delta -r^2 e^{-2\gamma} W^2\times\nonumber\\
 \cosh 2\delta
- 2r^2 UW\sinh 2\delta)du^2
 +2e^{2\beta}dudr
 +2r^2\times\nonumber\\(e^{2\gamma}U\cosh 2\delta
+W \sinh 2\delta)dud\theta
 +2r^2(e^{-2\gamma}W\times\nonumber\\
\cosh 2\delta + U\sinh
2\delta) \sin\theta dud\phi -r^2(e^{2\gamma}\cosh 2\delta d\theta^2\nonumber\\
+e^{-2\gamma}\cosh 2\delta \sin^2\theta d\phi^2
 +2\sinh 2\delta \sin\theta d\theta d\phi),\,\,\,\,\,\;\;\;\;\label{bsm}
\end{eqnarray}
where $\beta$,
$\gamma$, $\delta$, $U$, $W$, $V$ are functions of $x^0=u$,
$x^1=r$,
$x^2=\theta$, $x^3=\phi$. Observe that, unlike \cite{Vickers}, we
adopt the
signature $-2$ as in \cite{Burg}.

The general analysis of the field
equations is similar to  the one  in
\cite{Boal}, but of course expressions
are far more complicated (see
\cite{Sachs, Burg, Vickers} for details).
In
particular, there are now two news functions.

The asymptotic
expansion
of metric functions read in this case (see \cite{Burg} for
details)
\begin{eqnarray}
\gamma&=&cr^{-1} +
[C-c^3/6 -3cd^2/2]r^{-3} +...\;\;\;\, \label{26}
\end{eqnarray}
\begin{eqnarray}
\delta&=&dr^{-1} +
[H+c^2d/2 -d^3/6]r^{-3} + ...\;\;\;\,\label{27}
\end{eqnarray}
\begin{eqnarray}
U
&=&-\left(c_\theta + 2 c \cot{\theta} + d_\phi
\csc{\theta}\right)r^{-2}\nonumber\\
&+&
\left[2 N + 3 (c c_\theta + d d_\theta) + 4 (c^2 + d^3)
\cot{\theta}\right.\nonumber\\
& -& \left.2(c_\phi d - c d_\phi)\csc{\theta}\right]r^{-3} + \cdots
\label{U2}
\end{eqnarray}
\begin{eqnarray}
W  = - (d_\theta + 2 d
\cot{\theta} - c_\phi \csc{\theta} )
r^{-2}\;\;\;\;\;\;\;\;\;\;\;\;\;\;\;\;\;\;\;\;\;\;\;\;\;\;\nonumber\\
 + [ 2Q + 2(c_\theta d
- c d_\theta) + 3(c c_\phi + d
d_\phi)\csc{\theta}] r^{-3} + \cdots\;\;
\label{W}
\end{eqnarray}
\begin{eqnarray}
V= r - 2M - [N_\theta + \cot{\theta} + Q_\phi \csc{\theta}
- (c^2 + d^2)/2 \nonumber\\
- (c^2_\theta + d^2_\theta) - 4(c c_\theta +
d d_\theta)\cot{\theta}
- 4(c^2+ d^2)\cot^2{\theta}\nonumber\\
 - (c^2_\phi + d^2_\phi)\csc^2{\theta} +
4(c_\phi d - c
d_\phi)\csc{\theta}\cot{\theta}\;\;\;\;\;\;\;\;\;\;\;\,\nonumber \\
+ 2 (c_\phi
d_\theta - c_\theta d_\phi) \csc{\theta}] r^{-1} +
\cdots\;\;\;\;\;\;\;\;\;\;\;\;\;\;\;\;\;\;\;\;\;\;\;\;\;\;\;\;\;
\label{V2}
\end{eqnarray}
\begin{eqnarray}
\beta = - ( c^2+d^2) r^{-2}/4 + ...\;\;\;\;\;\;\;\;\;\;\;\;\;\;
\;\;\;\;\;\;\;\;\;\;\;\;\;\;\;\;\;\;\;\;\;\;\;
\label{be1}
\end{eqnarray}
where $c$, $C$, $d$, $H$, $N$, $Q$ and $M$ are now functions of $u$, $\theta$ and $\phi$. It can be shown \cite{Sachs, Burg,
Vickers} that field equations allow to
determine the $u$-derivatives of all
functions in $\beta$ and $\gamma$ with
the exceptions of news
functions
$c_u$ and
$d_u$, which remain arbitrary and whose existence represents a
clear--cut
criterium for the presence of gravitational radiation.

Let us
first calculate the vorticity for the congruence
of observers at rest in
(\ref{bsm}), whose four--velocity vector is given
by  $u^{\alpha} =A^{-1}\delta^{\alpha}_{u}$,
where now $A$ is given
by
\begin{eqnarray}
A&=&(Vr^{-1}e^{2\beta}-r^2 e^{2\gamma}U^2\cosh 2\delta\nonumber\\
&-&r^2 e^{-2\gamma} W^2
\cosh 2\delta -2r^2 UW\sinh
2\delta)^{1/2}.
\end{eqnarray}
Thus, (\ref{vv}) lead us to
\begin{equation}
\omega^\alpha=(\omega^u,\omega^r,\omega^\theta,\omega^\phi),
\end{equation}
where
\begin{eqnarray}
\omega^u&=&-\frac{1}{2A^2\sin\theta}
\{r^2e^{-2\beta}(WU_{r}-UW_{r})\nonumber \\
&+&\left[2r^2\sinh 2\delta
\cosh 2\delta (U^2e^{2\gamma}+W^2e^{-2\gamma}) \right.\nonumber\\
&+&\left. 4UWr^2\cosh^2 2\delta\right]e^{-2\beta}\gamma_{r}\nonumber \\
&+&2r^2e^{-2\beta}(W^2e^{-2\gamma}
-U^2e^{2\gamma})\delta_{r}\nonumber \\
&+&e^{2\beta}[e^{-2\beta}(U\sinh
2\delta+e^{-2\gamma} W\cosh
2\delta)]_{\theta}\nonumber \\
&-&e^{2\beta}[e^{-2\beta}(W\sinh
2\delta+e^{-2\gamma} U\cosh
2\delta]_{\phi}\}
\end{eqnarray}
\begin{eqnarray}
\omega^r&=&\frac{1}{e^{2\beta}\sin\theta}\{2r^2 A^{-2}[ (
(U^2e^{2\gamma}+W^2e^{-2\gamma})\sinh
2\delta \cosh 2\delta\nonumber \\
&+&UW\cosh^2 2\delta) \gamma_{u})+(W^2e^{-2\gamma}-U^2e^{2\gamma})\delta_{u}
+\frac{1}{2}(WU_{u}-UW_{u})]\nonumber \\
&+&A^2[A^{-2}(We^{-2\gamma}\cosh 2\delta+U\sinh 2\delta)]_{\theta} \nonumber \\
&-&A^2[A^{-2}(W\sinh 2\delta+Ue^{2\gamma}\cosh 2\delta)]_{\phi}\}
\end{eqnarray}
\begin{eqnarray}
\omega^\theta&=&\frac{1}{2
r^2\sin\theta}\{A^2e^{-2\beta}[r^2A^{-2}(U\sinh
2\delta+We^{-2\gamma}\cosh
2\delta)]_{r}\nonumber \\
 &-&e^{2\beta} A^{-2}[e^{-2\beta} r^2(U\sinh
2\delta+e^{-2\gamma} W\cosh
2\delta)]_{u}\nonumber \\
&+&e^{2\beta} A^{-2}(e^{-2\beta} A^2)_{\phi} \},
\end{eqnarray}
and
\begin{eqnarray}
\omega^\phi&=&\frac{1}{2r^2\sin\theta} \{
A^2e^{-2\beta}[r^2 A^{-2} (W\sinh 2\delta+Ue^{2\gamma}\cosh
2\delta)]_{r} \nonumber \\
&-&e^{2\beta} A^{-2}[r^2e^{-2\beta}(W\sinh
2\delta+Ue^{2\gamma}\cosh
2\delta)]_{u}\nonumber \\
&+&A^{-2}e^{2\beta}(A^2e^{-2\beta})_{\theta}\}.
\end{eqnarray}
Thus, for the leading term of
the absolute value of $\omega^\mu$ we
get
\begin{eqnarray}
\Omega&=&-\frac{1}{2r}\left[(c_{\theta u}+2c_{u}\cot\theta +
d_{\phi u} \csc\theta)^2\right.\nonumber\\
&+&\left.(d_{\theta u}+2d_{u}\cot\theta - c_{\phi u}
\csc\theta)^2\right]^{1/2}\label{On}
\end{eqnarray}
which of course reduces to
(\ref{Om2}) in the Bondi (axially and reflection
symmetric) case
($d=c_{\phi}=0$).

Next, calculation of the super--Poynting gives the
following
result
\begin{equation}
P_\mu=(0,P_r,P_\theta,P_\phi),
\end{equation}
where
the explicit terms are too long to be written at this point.

Although algebraic manipulation by hand is feasible for the Bondi
metric, for the Bondi--Sachs one is quite cumbersome. Thus, we
calculated the super--Poynting components using two different sets
of Maple scripts (available upon request). One, uses the Maple
intrinsic procedures to deal with tensors; the electric and magnetic
parts of the Weyl tensor were manipulated not explicitly up to the
leading term  output for the super--Poynting vector components. The
most important feature for this procedure, which leads to
considerable simplification, is the use of the shift vector $U^A=(U,W/\sin\theta)$
and the
2--surfaces metric of constant $u$
\begin{equation}
h_{AB}=\left (
\begin{array}{ll}

e^{2\gamma}\cosh{2\delta}&\,\,\,\sinh{2\delta}\sin\theta \\

\sinh{2\delta}\,\sin\theta &\,\,\,
e^{-2\gamma}\cosh{2\delta}\,\sin^2\theta
               \end{array}
\right ),
\end{equation}
where $A,\, B$ runs from $2$ to $3$. In our calculations we keep
these auxiliary variables, $U^A$ and $h_{AB}$, as far as was
possible. The other set of Maple scripts uses the GRTensor II
computer algebra package (running on Maple); in this case, all the
relevant objects (electric and magnetic parts of the Weyl tensor,
super--Poynting vector $P^\alpha$) were calculated exactly for the
metric (\ref{bsm}) performing next  a series expansion (at $r
\rightarrow \infty$) for the components of $P^\alpha$, taking into
account the expressions (\ref{26}--\ref{be1}) for the  metric functions
and keeping just the leading term in the series. Both approaches
lead us to the same results.

The leading terms for each super--Poynting component are
\begin{equation}
P_r=-2r^{-2}(d^2_{uu}+c^2_{uu}), \label{sp1}
\end{equation}
\begin{eqnarray}
P_\theta &=&-\frac{2}{r^{2}\sin\theta}\{
[2(d_{uu}^2+c_{uu}^2)c
+c_{uu}c_{u}+d_{uu}d_{u}]\cos\theta \nonumber
\\
&+&\left[c_{uu}c_{\theta
u}+d_{uu}d_{\theta
u}+(c_{uu}^2+d_{uu}^2)c_{\theta}\right]\sin\theta
\nonumber \\
&+&c_{uu}d_{\phi u}-d_{uu}c_{\phi u}
+(d_{uu}^2+c_{uu}^2)d_{\phi}\}, \label{sp2}
\end{eqnarray}
\begin{eqnarray}
P_\phi&=&\frac
{2}{r^2}\{ 2[c_{uu}d_{u}-d_{uu}c_{u}-(d_{uu}^2+c_{uu}^2)d]\cos\theta
\nonumber \\
&+&\left[c_{uu}d_{\theta
u}-d_{uu}c_{\theta
u}-(c_{uu}^2+d_{uu}^2)d_{\theta}\right]\sin\theta
+\nonumber \\
&+&(d_{uu}^2+c_{uu}^2)c_{\phi}-(c_{uu}c_{\phi u}+d_{uu}d_{\phi
u}) \}. \label{sp3}
\end{eqnarray}
from which it follows
\begin{eqnarray}
P^{\phi} &=&
-\frac{2}{r^4 \sin^2\theta}\{ \sin\theta \left[ d_{u\theta}
c_{uu} - d_{uu}
c_{u \theta} \right]+2\cos\theta\times\nonumber\\
&& \left[c_{uu} d_{u} -
d_{uu} c_{u}
\right]- [c_{uu} c_{u\phi}+ d_{uu}d_{u \phi}]\}
.\label{fi}
\end{eqnarray}
which of course vanishes in the Bondi case.
\section{Discussion}

We have seen so far that gravitational radiation produces vorticity in the congruence of observers at rest in the frame of (\ref{bsm}). We conjecture that such vorticity is
caused by the presence of  a flow of super--energy, as described by the super--Poynting vector (\ref{p1}).

In the case of axial and reflection symmetry (Bondi), the plane of rotation is
orthogonal to the unit vector $\hat{e}_\phi$, in the $\phi$ direction, and accordingly it is the $P^\theta$ component the responsible for such vorticity, whereas the azimuthal component
$P^{\phi}$ vanishes, in agreement with the fact that the symmetry of the Bondi metric excludes rotation along $\phi$ direction. 

In the general case, we have shown that the vorticity vector has also components along  $\hat{e}_r$ and  $\hat{e}_\theta$, implying thereby rotations on the corresponding orthogonal
planes. In particular,  rotations in the
$\phi$ direction are now allowed ($\omega^{\theta}\neq 0$), and therefore we should expect a  non--vanishing $P^\phi$, which is in fact what happens.

Thus, we have shown that  there is always a non--vanishing component of $P^\mu$, on the plane orthogonal to a unit vector along which there is a non--vanishing component of
vorticity. Inversely, $P^\mu$ vanishes on a plane orthogonal to a unit vector along which the component of
vorticity vector vanishes (Bondi). This  supports further our conjecture about the link between the super--Poynting vector and vorticity.

\section*{Acknowledgments}
JC gratefully acknowledges financial support from the
Spanish Ministerio de Educaci\'on y Ciencia through the grant 
FPA2004-03666.  This work was supported partially by FONACIT
under grants S1--98003270 and  F2002000426; by CDCHT--ULA under grant C--1267--04--05--A.
LH wishes to thank Roy Maartens for pointing out the super--Poynting vector as defined in \cite{Basset}.
WB is grateful for the hospitality received at the Pittsburgh Supercomputing Center, where this work was finished.


\thebibliography{99}
\bibitem{HHP} L  Herrera and J L
Hern\'andez-Pastora, {\it Class. Quantum.
Grav.} {\bf 17} 3617
(2000)
\bibitem{Valiente} J Valiente  {\it Class. Quantum. Grav.} {\bf 18}
4311
(2001)
\bibitem{HSC} L Herrera, N O Santos and J Carot {\it J. Math.
Phys} {\bf 47
} 052502 (2006)
\bibitem{Felice} F Sorge, D Bini, F de Felice
{\it Class. Quantum Grav.}
{\bf 18} 2945 (2001)
\bibitem{bonleter} W B
Bonnor {\it Phys. Lett. A} {\bf 158} 23 (1991)
\bibitem{Feymann} R P
Feynman, R B Leighton and M Sand {\it Lectures on
Physics} {\bf II}
(Addison--Wesley, Reading, 1964) pp. 27, 28
\bibitem{HCol} L Herrera, G A
Gonz\'alez, L A Pach\'on and J A Rueda {\it
Class. Quantum Grav.} {\bf 23} 2395
(2006).
\bibitem{Bel} L Bel {\it C. R. Acad. Sci.} {247} 1094 (1958);
{\it Cah.
de Phys.} {\bf 16} 59 (1962);  {\it Gen. Rel.
Grav.} {\bf 32}
2047 (2000)
\bibitem{Basset} R Maartens and B A Basset {\it Class.
Quantum Grav.} {\bf
15} 705 (1998).
\bibitem{Boal} H Bondi, M G J  van der
Burg and A W K  Metzner
{\it Proc. R. Soc.} {\bf A269} 21
(1962)
\bibitem{Sachs} R Sachs
{\it Proc. R. Soc.} {\bf A270} 103
(1962)
\bibitem{Burg} M G J van der Burg,
{\it Proc. R. Soc.} {\bf A294}
112 (1966)
\bibitem{Vickers} R A D'Inverno and J A Vickers {\it Phys. Rev.
D } {\bf
56} 772 (1997).
\bibitem{Rindler} W Rindler and V Perlick {\it
Gen. Rel. Grav.} {\bf22}
1067 (1999)
\end{document}